\documentclass{article}
\usepackage{amsfonts}
\usepackage{amsthm}
\usepackage{amssymb}

\newlength{\defbaselineskip}
\newcommand{\setlinespacing}[1]%
           {\setlength{\baselineskip}{#1 \defbaselineskip}}

\newcommand{\singlespacing}{\setlength{\baselineskip}{\defbaselineskip}}
\newcommand{\onehalfspacing}{\setlength{\baselineskip}{1.5 \defbaselineskip}}

\theoremstyle{plain}
\newtheorem{thm}{Theorem}[section]

\theoremstyle{definition}
\newtheorem{defn}[thm]{Definition}
\theoremstyle{remark}
\newtheorem{rem}[thm]{Remark}

\def\BA{\mathcal A}
\def\BB{\mathcal B}
\def\KB{\mathfrak B}

\def\BO{\mathcal O}
\def\BS{\mathcal S}

\begin{document}
\begin{center}
{\LARGE\textbf{On the notion of reconstruction of}} \vskip 0.125
true in {\LARGE\textbf{quantum theory}}
      \vskip 0.3 true in
{\large \bf Alexei Grinbaum}
\vskip 0.15 true in
\par
{\it LPHS -- Archives Henri Poincar\'e (CNRS UMR 7117), \\ 23 bd
Albert Ier, 54015 Nancy, France
\par Email alexei.grinbaum@polytechnique.edu}
\par\vskip 0.2 true in
\end{center}
\begin{abstract}\noindent
What belongs to quantum theory is no more than what is needed for
its derivation. We argue for an approach focusing on reconstruction
rather than interpretation of quantum mechanics. After discussing
the concept of reconstruction, we analyze the problem of
metatheoretic justification of the choice of axioms and then study
several examples of reconstruction.
\end{abstract}

\onehalfspacing

\section{What is wrong with interpreting quantum mechanics?}

Ever since the first days of quantum mechanics physicists as well
as philosophers tried to interpret it, understanding this task as
a problem of giving to the new, puzzling physical theory a clear
meaning. Looking globally, this enterprize failed: still today we
have no consensus on what the meaning of quantum theory \emph{is}.
Proposed answers are many but none of them has won overall
recognition. Perhaps the most remarkable manifestation of the
failure to interpret quantum mechanics is the attitude taught to
most young physicists in lecture rooms and research laboratories
in the last half century, ``Write down equations and calculate! No
need to ask questions!''

Why did attempts at univocal interpretation fail? Many answers are
possible, and among them we favor two, both showing that there is
an intrinsic deficiency in the idea of interpreting
philosophically a physical theory.

The first answer is that to a physical theory one would naturally
like to give a \emph{physical} meaning in the Greek sense of
$\varphi\acute{\upsilon}\sigma\iota\varsigma$, i.e. we -- as part
of the physicists' audience -- expect to be told a story about the
immanent, fundamental nature. This is because we casually tend to
apply physical theory to the phenomenal world to learn something
about the latter, and not the world to physical theory in order to
invent a meaning of the theory. Physical theory is above all a
tool for predicting the yet unobserved phenomena; so employing the
existing knowledge and experience of the world to interpreting
physics runs counter to its basic function as a scientific theory.
However, notwithstanding such an against-the-grain direction in
which a philosophical interpretation operates, the former does not
necessarily lead to a formal contradiction that would invalidate
the interpretation logically; more modestly but perhaps no less
irritatingly, at the end one is often left with a feeling of being
excluded from the mainstream research. Further, as the physics of
today is inseparable from mathematics, a meaning cannot be
physical and thus satisfactory if it is merely heaped over and
above the mathematical formalism of quantum mechanics, instead of
coming all the way along with the formalism as it rises in a
derivation of the theory.

The second answer is that we live in a situation where objective
truth has been appropriated by science, and to pass public
ratification every increase in knowledge must confront experimental
setups. In this world an interpretation can only then be considered
satisfactory when it becomes an integral part of science. This is
not unprecedented in the history of ideas: indeed, many
philosophical questions with the advent of empirical science ceased
to be perceived as philosophical and are now treated as scientific.
Of the problem of interpreting quantum mechanics, as much as
possible must be moved into the area of the scientific; only then
will the puzzle disappear.

\section{Reconstruction of physical theory}

\subsection{Schema}

We call \emph{reconstruction} a following schema: Theorems and
major results of physical theory are formally derived from simpler
mathematical assumptions; these assumptions or axioms, in turn,
appear as a representation in the formal language, of a set of
physical principles. Thus, reconstruction consists of three parts:
a set of physical principles, their mathematical representation,
and a derivation of the formalism of the theory.

Contrary to an interpretation, reconstruction of physical theory
acquires supplementary persuasive power which arises from the use
of mathematical derivation. Established as a valid mathematical
result, the theorems and equations of the theory become
unquestionable and free of suspicion. 'Why is it so?'---'Because
we derived it.' The question of meaning, previously asked with
regard to the formalism, is removed and now bears only on the
selection of the principles. No room for mystery remains in what
concerns the meaning of the theory's mathematical apparatus.

\subsection{Selection of first principles}

That who wishes to attempt a reconstruction of physical theory
must formulate the foundational principles which he or she
believes plausible and translate them into mathematical axioms.
Then the rest of the theory will be constructed ``mechanically,''
by means of a formal derivation. The choice of axioms must be the
only allowed freedom in the whole construction. It is commonplace
to say that it is not easy to exhibit an axiomatic system that
would stand to such requirements, especially in the case of
quantum theory.

First, how does one judge which axioms are plausible? Prior to
pronouncing a judgment, one must develop an intuition of what is
plausible about quantum theory and what is not. This can be only
achieved by \textit{practicing} the theory, i.e. by taking its
prescriptions at face value, applying them to systems under
consideration in particular tasks, and obtaining results. In
short, one needs to acquire a real ``know-how'' above and beyond
the theoretical knowledge that quantum mechanics could solve such
and such problems. Intuition then develops from experience; it
cannot arise from abstract knowledge ``in principle.'' However, it
is important to say that taking the prescriptions of quantum
theory at face value, applying them and obtaining results will not
yet make things \emph{clear} about quantum mechanics. One can
possess the knowledge about how to apply a certain tool without
caring about the structure of the tool or its meaning. The quantum
mechanical know-how purely serves as such tool for developing
one's intuition about which idea is plausible and can become a
foundational principle and which candidate idea will not pass the
test.

Second, what shall one require from the first principles? As we
stated above, they must be simple \emph{physical statements} whose
meaning is immediately, easily accessible to a scientist's
understanding. They must also be such as to permit a clear and
unambiguous translation of themselves into mathematically formulated
axioms. A derivation of quantum theory will then rely on these
axioms.

\subsection{Status of first principles}

Reconstruction program includes a \emph{derivation} of quantum
theory, but in the previous section one was told to \emph{apply}
and \emph{use} it in order to motivate the derivation. Is there a
logical vicious circle here? We submit that there is none, and
this thanks to the status of first principles. Namely, they should
not necessarily be viewed as ultimate truths about nature.
Independently of one's ontological commitments, the first
principles only have a minimal epistemic status of being
postulated for the purpose of reconstructing the theory in
question. Like in the 19th-century mathematics, in theoretical
physics the axiomatic method is to be separated from the attitude
which yet the Greeks had toward axioms: that they represent the
truth about reality. Much of the progress of mathematics is due to
understanding that an axiom can no longer be considered ultimate
truth, but merely a basic structural element, i.e. assumption that
lies in the foundation of a certain theoretical structure. In
mathematics, after departing from the Greek concept of axiom,
``not only geometry, but many other, even very abstract, theories
have been axiomatized, and the axiomatic method has become a
powerful tool for mathematical research, as well as a means of
organizing the immense field of mathematical knowledge which
thereby can be made more surveyable''~\cite{heyting}. A similar
attitude is to be taken with respect to axioms used for a formal
derivation of physical theory. A short prescription would sound
something like this, ``If the theory does not tell you that the
states of the system are ontic states, do not take them to be
ontic.''

To explain the above prescription, return first to the idea that, in
developing an intuition with respect to the plausibility of the
foundational principles used to derive a theory, one takes this
theory for a given and applies it practically so as to acquire a
know-how that would justify the choice of principles. Now, when one
is working with several physical theories, ideas that have
previously been used as foundational in theory I, may turn out to be
derivative (i.e., theorems) in theory II. A good example comes from
the case of thermodynamics and statistical physics, or macroscale
hydrodynamics and low-level molecular theory of liquids. Such
considerations show the limits of philosophical assumptions that one
can make about the status of first principles used in reconstruction
of a given physical theory. Indeed, generically nothing can be said
about their ontological content or the ontic commitments that arise
from the principles. It is more economical and would amount to a
certain ``epistemological modesty'' to treat the foundational
principles as axioms \emph{hic et nunc}, i.e. in a given theoretical
description. Having taken this position, ask then the reconstructed
theory itself if it allows a realist point of view or imposes limits
on it; and while in general the status of first principles as
ultimate truths about reality is not a necessity, certain
reconstruction programs are such that this status can be safely, or
almost, attributed to the principles within a particular
reconstruction in question.

Reconstruction of physical theory has its main advantage compared
to philosophical interpretation of the theory in that it moves a
number of questions, previously being thought of as philosophical,
to the realm of science, and this in virtue of the mathematical
derivation which the reconstruction program operates. However,
philosophical problems do not altogether disappear; they still
apply to the first principles and take the form of a problem of
their \emph{justification}. Evidently, it is a minimal logical
condition that such a justification should not be seen as a
mathematical deduction of the principles from the theory in whose
very foundations they lie. The task of justification is external
to the reconstruction program and must be executed by the one with
a different set of presuppositions, i.e. by taking the theory for
a given and motivating from there why the principles in question
are simple, physical, and plausible. Therefore, philosophy is not
fully chased out of physics. On the contrary, by demarcating the
frontier between what can be treated as a scientific question and
what belongs to metatheory, one contributes to a better
understanding of the structure of the theory and of those of its
foundational postulates which require a metatheoretic
interpretation and justification.

\section{Examples of reconstruction}

\subsection{Early examples}

A particularly well-known example of reconstruction is the special
theory of relativity. Since Einstein's 1905 work special
relativity is a well-understood physical theory; but it is equally
well-known that its formal content, i.e. Lorentz transformations,
were written by Lorentz and Poincar\'{e} and not by Einstein, and
this several years before 1905. Lorentz transformations were
fiercely debated and many interpretations of what they mean were
offered, and among them quite a plausible one about interactions
between bodies and the ether. However, when Einstein came, things
suddenly became clear and the debate stopped. This was because
Einstein gave a few simple physical principles from which he
\textit{derived} Lorentz transformations, therefore closing the
attempts to append philosophy and give a meaning \textit{a
posteriori}, to an already working formalism. Einstein's idea was
to assume that there is no absolute, but only a relative, notion
of simultaneity and that the velocity of light is constant. Once a
derivation starting from these principles has been taken through,
the physical meaning of Lorentz transformations stood clear and
special relativity has not raised any controversy ever since.

Einstein's reconstruction of special relativity is an example of
theory where the first principles are understood as truths about
reality. That the speed of light is constant and that there is no
absolute notion of simultaneity is now routinely taken to be
objectively true and well-established facts about nature. Thus, we
have here a case in which, although in general it is not a
necessity, the first principles do acquire a particular
ontological status.

Moving away from special relativity, we submit that reconstruction
is the exclusive way to make things \textit{clear} about quantum
mechanics. As such, this idea is not novel but has been in the air
for some time, and a concise statement can for example be found in
Rovelli \cite{RovRQM},
\begin{quote}
Quantum mechanics will cease to look puzzling only when we will be
able to \textit{derive} the formalism of the theory from a set of
simple physical assertions (``postulates,'' ``principles'') about
the world. Therefore, we should not try to append a reasonable
interpretation to the quantum mechanical formalism, but rather to
\textit{derive} the formalism from a set of experimentally
motivated postulates.
\end{quote}

What is interesting is that in the last decade reconstruction
became a major trend in the foundations of quantum mechanics. But
before describing this recent work let us first look further back
in the history of quantum mechanics: there too axiomatic
derivations occupy an eminent place. The first paper where quantum
mechanics was treated axiomatically appeared shortly after the
creation of quantum mechanics itself: in 1927 Hilbert, von Neumann
and Nordheim stated their view of quantum mechanics as the one in
which ``(the theory's) analytical apparatus, and the arithmetic
quantities occurring in it, receives \textit{on the basis of the
physical postulates} a physical interpretation. Here, the aim is
to formulate the physical requirements so completely that the
analytical apparatus is just uniquely determined. Thus the route
is of axiomatization''~\cite{HvNN}. It is on this route of
axiomatization that von Neumann in collaboration with Birkhoff was
led to study the logic of quantum mechanics \cite{bvn}. Following
their work, many axiomatic systems were proposed, e.g. Zieler
\cite{zieler}, Varadarajan \cite{varada,varada2}, Piron
\cite{piron,piron72}, Kochen and Specker \cite{kochspeck65},
Guenin \cite{guenin}, Gunson \cite{Gunson}, Jauch \cite{jauch},
Pool \cite{pool,pool2}, Plymen \cite{plymen}, Marlow
\cite{marlow}, Beltrametti and Casinelli \cite{bc}, Holland
\cite{holland}, or Ludwig \cite{ludwig}. Another branch of
axiomatic quantum theory, the algebraic approach was first
conceived by Jordan, von Neumann and Wigner \cite{jvNw} and later
developed by Segal \cite{segal1,segal2}, Haag and Kastler
\cite{hk}, Plymen \cite{plymen2}, Emch \cite{Emch} and others; for
a recent review, see~\cite{Freden}.

However, a vast majority of these axiomatic derivations do not
fall under our notion of reconstruction, as they were based on
highly abstract mathematical assumptions and not, as we required,
on simple physical principles. Consider for instance the exemplary
work by Mackey \cite{mackey57,mackey63}.

Mackey develops quantum mechanics as follows. Take a set $\KB$ of
all Borel subsets of the real line and suppose we are given two
abstract sets $\BO$ (a to-be space of observables) and $\BS$ (a
to-be space of states) and a (to-be probability) function $p$
which assigns a real number $0\leq p(x,f,M)\leq 1$ to each triple
$x,f,M$, where $x$ is in $\BO$, $f$ is in $\BS$, and $M$ is in
$\KB$. Assume certain properties of $p$ listed in axioms M1-M9:
\begin{description}
    \item[M1] Function $p$ is a probability measure. Mathematically,
    we have $p(x,f,\emptyset)=0$, $p(x,f,\mathbb{R})=1$, and $p(x,f,M_1 \cup M_2
\cup M_3 \ldots)=\sum _{n=1} ^{\infty} p(x,f,M_n)$ whenever the
$M_n$ are Borel sets that are disjoint in pairs.

\item[M2] Two states, in order to be different, must assign
different probability distributions to at least one observable;
and two observables, in order to be different, must have different
probability distributions in at least one state. Mathematically,
if $p(x,f,M)=p(x^\prime,f,M)$ for all $f$ in $\BS$ and all $M$ in
$\KB$ then $x=x^\prime$; and if $p(x,f,M)=p(x,f^\prime,M)$ for all
$x$ in $\BO$ and all $M$ in $\KB$ then $f=f^\prime.$

\item[M3] Let $x$ be any member of $\BO$ and let $u$ be any real
bounded Borel function on the real line. Then there exists $y$ in
$\BO$ such that $p(y,f,M)=p(x,f,u^{-1}(M))$ for all $f$ in $\BS$
and all $M$ in $\KB$.

\item[M4] If $f_1$, $f_2$,\ldots are members of $\BS$ and
$\lambda_1+\lambda_2+\ldots =1$ where $0\leq \lambda_n \leq 1,$
then there exists $f$ in $\BS$ such that $p(x,f,M)=\sum ^{\infty}
_{n=1} \lambda_n p(x,f_n,M)$ for all $x$ in $\BO$ and $M$ in
$\KB.$

\item[M5] Call \textit{question} an observable $e$ in $\BO$ such that
$p(e,f,\{0,1\})=1$ for all $f$ in $\BS$. Questions $e$ and
$e^\prime$ are disjoint if $e\leq 1-e^\prime.$ Then a question
$\sum ^{\infty} _{n=1} e_n$ exists for any sequence $(e_n)$ of
questions such that $e_m$ and $e_n$ are disjoint whenever $n\neq
m.$

\item[M6] If $E$ is any compact, question-valued measure then
there exists an observable $x$ in $\BO$ such that $\chi _M (E) =
E(M)$ for all $M$ in $\KB$, where $\chi _M$ is a characteristic
function of $M.$

\item[M7] The partially ordered set of all questions in quantum
mechanics is isomorphic to the partially ordered set of all closed
subspaces of a separable, infinite-dimensional Hilbert space.

\item[M8] If $e$ is any question different from 0 then there
exists a state $f$ in $\mathcal{S}$ such that $m_f(e)=1$.

\item[M9] For each sequence $(f_n)$ of members of $\BS$ and each
sequence $(\lambda _n)$ of non-negative real numbers whose sum is
1, one-parameter time evolution group $V_t:\BS\mapsto\BS$ acts as
follows: $V_t \left( \sum ^{\infty} _{n=1} \lambda_n
f_n\right)=\sum ^{\infty} _{n=1} \lambda_n V_t(f_n)$ for all
$t\geq 0;$ and for all $x$ in $\BO$, $f$ in $\BS$, and $M$ in
$\KB$, $t\rightarrow p(x,V_t(f),M)$ is continuous.
\end{description}

In Mackey's nine axioms all essential features of the quantum
formalism are directly postulated in their mathematical form: the
Hilbert space structure in M5-M8, state space and the probabilistic
interpretation in M1-M4, and time evolution in M9. It is not at all
clear where these mathematical definitions come from and how one
justifies them on physical rather than formal grounds. In fact,
Mackey's concern in the early 1950s was with a precise mathematical
axiomatization of quantum mechanics rather than with the question of
what quantum mechanics tells us about the world and then
reconstructing its formalism from a set of such fundamental ideas.
Thus, stage 1 of reconstruction, at which one formulates physical
principles, is absent from Mackey's work, and instead one starts
directly at stage 2 where axioms appears as formal, mathematical
definitions.

Later, Mackey's axioms M5-M8 were reformulated in the language of
quantum logic, thereby rephrasing the assumptions that underlie the
Hilbert space structure. This was the case, most prominently, in
\cite{jauch,piron,piron72} and also in a seminal book \cite{bc}.
Quantum logical assumptions are simple enough to be accessible for
direct comprehension, in contrast to Mackey's mathematically
formulated axioms, but they tend to be linguistic rather than
physical. This means that one typically argues that it makes no
sense to speak about certain terms other than if some suitable
``trivial'' properties had been postulated, e.g. the notion of
proposition is only meaningful if like in Ref.~\cite{demopou}
negation or partial order, or like in Ref.~\cite{bc} implication,
are defined. Although we fully acknowledge that linguistic \emph{a
priori} arguments can be interesting and powerful, we however
separate them from the reconstruction program as introduced above:
in the latter, first principles from which the theory is derived
should have a \emph{physical} meaning, i.e. tell us something
directly and intuitively apprehensible about the world and quantum
theory as describing our knowledge of it. Such principles, ideally,
should be independent of a particular formalism in which we derive
quantum theory and therefore should not rely on the language of
quantum logic as just one among many such formalisms.

\subsection{Contemporary examples}

Among the modern developments, an interesting example of
reconstruction comes from the instrumentalist derivation of
quantum theory from ``five reasonable axioms'' by
Hardy~\cite{hardy}. Hardy's ``reasonable axioms'' set up a link
between two quantities, $K$ and $N$, which play a fundamental role
in the reconstruction. $K$ is the number of degrees of freedom of
the system and is defined as the minimum number of probability
measurements needed to determine the state. Dimension $N$ is
defined as the maximum number of states that can be reliably
distinguished from one another in a single measurement. The axioms
then are:
\begin{description}
    \item[H1] \textit{Probabilities}. Relative frequencies (measured
    by taking the proportion of times a particular outcome is
    observed) tend to the same value for any case where a given
    measurement is performed on an ensemble of $n$ systems prepared
    by some given preparation in the limit as $n$ becomes infinite.
    \item[H2] \textit{Simplicity}. $K$ is determined by a function
    of $N$ where $N=1,2,\ldots$ and where, for each given $N$, $K$
    takes the minimum value consistent with the axioms.
    \item[H3] \textit{Subspaces}. A system whose state is
    constrained to belong to an $M$ dimensional subspace behaves
    like a system of dimension $M$.
    \item[H4] \textit{Composite systems}. A composite system
    consisting of subsystems $A$ and $B$ satisfies $N=N_A N_B$ and
    $K=K_A K_B.$
    \item[H5] \textit{Continuity}. There exists a continuous
    reversible transformation on a system between any two pure
    states of that system.
\end{description}

Although four of the H1-H5 axioms use mathematical language in the
formulation, their meaning in Hardy's instrumentalist setting can
be grasped much easier than the meaning of Mackey's axioms M1-M9.
In fact, this meaning is already suggested by the names given to
the axioms by Hardy. Therefore H1-H4 can be rephrased into
\emph{physical principles} from which one derives the formalism of
the theory and thus provide an example of reconstruction. None of
these principles is trivial: for H1, assume that probability
introduced instrumentally as relative frequency of measurements is
a well-defined concept and obeys the laws of probability theory;
for H2, assume that the number of parameters needed to
characterize a state is directly linked to the number of states
that can be distinguished in one measurement; for H3, that the
linear structure of state space shrinks accordingly to the maximum
number of states of the system distinguishable in one measurement;
for H4, assume multiplicability of the quantity defined as
dimension and of the quantity defined as the number of degrees of
freedom. Now formulate these assumptions mathematically and use
Hardy's theorems to derive from them the full-blown formalism of
quantum mechanics. A particular instrumental philosophy does not
play a crucial role in the derivation: Hardy himself acknowledges
that his axioms can be adopted by a realist as well as a hidden
variable theorist or a partisan of collapse interpretations. Thus,
the choice of underlying philosophy is not critical for
derivation, and Hardy's reconstruction advances our understanding
of quantum theory irrespectively of the justification which one
may have for the axioms. What matters are the simple physical
principles formulated as axioms H1-H4. We shall see an opposite
example in the next section, in which the justification used for
fundamental principles will limit the area in which operates the
mathematical derivation.

Still, it is not so clear whether axiom H5 has a \emph{physical}
meaning. Because it is this axiom that makes the theory quantum
rather than classical, the reconstruction program cannot be said to
be fully implemented and taken to its logical conclusion. To further
illustrate this point, we distinguish two types of continuity
assumptions that are made in axiomatic derivations of quantum
theory. Continuity assumptions of type 1 select the correct type of
numeric field which is used in the construction of the Hilbert space
of the theory; namely, of the field $\mathbb{C}$ of complex numbers.
Sol\`er's theorem~\cite{soler} or Zieler's axioms~\cite{zieler} are
examples of type 1 continuity assumptions. Hardy's case is different
and is an example of the continuity assumptions of type 2, which are
made in order to bring in the superposition principle. Other such
assumptions include Gleason's non-contextuality~\cite{gleason},
Brukner's and Zeilinger's homogeneity of parameter space~\cite{BZ},
Landsman's two-sphere property~\cite{land}, and Holland's axioms C
and D~\cite{holland} which bear a particular resemblance to Hardy's
H5: \begin{quote}\begin{description}
\item[(C)]Superposition principle for pure states:
\begin{description}
  \item[1.] Given two different pure states (atoms) $a$ and $b$, there is
at least one other pure state $c$, $c\neq a$ and $c\neq b$ that is a
superposition of $a$ and $b$.
  \item[2.] If the pure state $c$ is a superposition of the distinct
pure states $a$ and $b$, then $a$ is a superposition of $b$ and $c$.
\end{description}
\item[(D)] Ample unitary group: Given any two orthogonal pure states $a,b\in \mathcal{L}$,
there is a unitary operator $U$ such that $U(a)=b$.
\end{description}\end{quote}
We see that various axiomatic systems of quantum theory contain,
under one form or another, the assumption of continuity and it is
this assumption which is largely responsible for making things
quantum. Whatever the framework of the reconstruction is, bringing
in topological considerations is essential. As it is exceedingly
difficult to formulate a physical principle which may give a
meaning to the continuity assumption of type 2, all reconstruction
programs suffer here from the intrusion of an element of
mathematical abstraction.

The above critique concerning the continuity axiom applies to
another examples of reconstruction initially proposed by
Rovelli~\cite{RovRQM} and that we developed
elsewhere~\cite{grinbijqi,grinbfpl}. Here, the reconstruction starts
from two information-theoretic axioms:
\begin{description}
    \item[R1] There exists a maximum amount of relevant information that can be
extracted from a system.
    \item[R2] It is always possible to obtain new information
    about the system.
\end{description}
From these axioms and with the help of supplementary mathematical
assumptions one derives the formalism of quantum mechanics. While
the supplementary assumptions cast a shadow on the conceptual
clarity of the reconstruction much in the same fashion as does H5,
the whole program presents itself differently from Hardy's
instrumentalism. The mathematical derivation being still devoid of
ontological commitments, justification of the first principles
which we propose cannot refer to an ontology, except for an
arguably problematic case in which one would be prepared to take
information for a fundamental building block of reality. Rather,
by reconstructing quantum theory from information-theoretic
principles we point at its epistemological character and at its
role as a theory of (a certain kind of) knowledge; i.e. with
certain limits being imposed on the kind of information one may be
dealing with, the most general theory of this information takes
the form of quantum theory. Here again reconstruction appears more
appealing than a mere interpretation as it leaves room for any
justification of the first principles different from ours. Indeed,
one may wish to adopt an ontological picture to justify R1-R2 or
take no position at all with respect to ontology. At the same
time, regardless of a specific philosophical justification of
first principles, the meaning of quantum theory stands clear:
quantum theory is a general theory of information constrained by
certain information-theoretic principles.

\subsection{CBH reconstruction}\label{cbhsect}

Clifton, Bub and Halvorson (CBH) offer an example of a set of
quantum informational postulates from which one derives the
structure of quantum theory~\cite{Bub}. CBH postulate three
fundamental principles:
\begin{description}
\item[CBH1] No superluminal information transfer via measurement.
\item[CBH2] No broadcasting. \item[CBH3] No bit
commitment.\end{description}

To give a mathematical formulation of these principles, CBH use
the $C^*$-algebraic formalism. Consider a composite system,
$\BA+\BB$, consisting of two component subsystems, $\BA$ and
$\BB$, understood as $C^*$-algebras.

\begin{defn} Operation $T$ on algebra $\BA\vee\BB$ conveys
no information to Bob if
\begin{equation}(T^{*}\rho) |_{\BB}=\rho |_{\BB}\mathrm{\;for\;all\;states\;}
\rho\mathrm{\;of\;}\BB.\label{defnoinf}\end{equation}\label{defnoinff}\end{defn}
An operation here is understood as a completely positive linear map
on an algebra, and $T^{*}\rho$ is a state over the algebra defined
for every state $\rho$ on the \textit{same} algebra as
\begin{equation}
(T^{*}\rho)(A)=\frac{\rho(T(A))}{\rho(T(I))}\label{eq84}
\end{equation}
at the condition that $\rho(T(I))\neq 0$. Nonselective
measurements $T$ are the ones that have $T(I)=I$, and then
$\rho(T(I))=\rho(I)=||\rho ||=1$. CBH explain that, in their view,
Definition~\ref{defnoinff} entails
\begin{equation}T(B)=B\mathrm{\;for\;all\;}B\in\BB.\label{cninf}\end{equation}

CBH then assert that if the condition (\ref{cninf}) holds for all
self-adjoint $B\in\BB$ and for all $T$ of the form
\begin{equation}T=T_E(A)=E^{1/2}AE^{1/2}+(I-E)^{1/2}A(I-E)^{1/2},\label{teform}\end{equation}
where $A\in \BA\vee\BB$ with $\BA$ and $\BB$ being
$C^*$-independent, and $E$ is a positive operator in $\BA$, then
algebras $\BA$ and $\BB$ are kinematically independent, i.e. all
$A\in\BA$ and $B\in\BB$ commute \cite[Theorem~1]{Bub}. Thus
kinematic independence is derived from the assumption of
$C^*$-independence and from the condition (\ref{defnoinf}), where
$C^*$-independence is brought into the discussion to grasp the
meaning of the fact that systems $\BA$ and $\BB$ are distinct.
Mathematically, $C^*$-independence means that for any state $\rho_1$
over $\BA$ and for any state $\rho_2$ over $\BB$ there is a state
$\rho$ over $\BA\vee\BB$ such that $\rho | _{\BA}=\rho_1$ and $\rho
| _{\BB}=\rho_2$. As for Definition~\ref{defnoinff}, the authors
take it to be a mathematical representation of Axiom CBH1.

According to the authors, the meaning of CBH1 is that when Alice
and Bob perform local measurements, Alice's measurements can have
no influence on the statistics for the outcomes of Bob's
measurements, and vice versa. CBH also say that ``otherwise this
would mean instantaneous information transfer between Alice and
Bob'' and ``the mere performance of a local measurement (in the
nonselective sense) cannot, in and of itself, transfer information
to a physically distinct system.'' Upon reading these statements,
one has a feeling that for CBH \textit{distinct} and
\textit{distant} are synonyms. This identification of terms might
indeed be a tacit assumption among quantum information theorists
who do not have to worry about relativistic effects, but in the
full-blown $C^*$-algebraic framework, as well as in the general
philosophical context, meaning of the two words is certainly
different. We have here an example showing how the initial quantum
informational language of the fundamental principles CBH1-CBH3
constrains the use of the algebraic formalism to situations where
fundamental principles make sense from the point of view of
quantum information, while in fact the formalism could also be
used in other, more complex situations. Unlike Hardy's derivation
which was independent of the particular instrumental justification
of its fundamental principles, the CBH reconstruction cannot be
taken through outside the field of quantum information, because
its mathematics, while being still valid outside this field, will
require additional justification. Apart from the identification of
terms `distant' and `distinct,' such is also the case with time
evolution, which is tacitly taken by CBH to be the usual quantum
mechanical time evolution, while in the general $C^*$-algebraic
framework this is not at all the case and a variety of different
``temporal'' evolutions are possible. One then avoids this problem
at the price of confining oneself to the quantum informational
paradigm.

Equating Definition~\ref{defnoinff} with Axiom CBH1 requires
particular attention to the mathematical details, and a point has
to be made about CBH's proof. If, following the authors, in this
definition $\rho$ is to be taken as a state over $\BB$, then the
definition does not make sense: operation $T$ is defined on
$\BA\vee\BB$ and consequently, in accordance with (\ref{eq84}),
$T^{*}\rho$ is defined for the states $\rho$ over $\BA\vee\BB$. If
one follows the CBH definition with a state $\rho$ over $\BB$,
then there would be no need to write $\rho |_{\BB}$ as CBH do, for
a simple reason that $\rho |_{\BB} = \rho$. To suggest a remedy,
we extend the reasoning behind this definition and reformulate it
in three alternative ways.
\begin{itemize}\item The first one is to require that in
Definition~\ref{defnoinff} the state $\rho$ be a state over the
algebra $\BA\vee\BB$. \item The second alternative is to consider
states $\rho$ on $\BB$ but to require a different formula, namely
that $(T|_{\BB})^{*}\rho=\rho$ as states over $\BB$. \item
Finally, the third alternative proceeds as follows: Take arbitrary
states $\rho _1$ over $\BA$ and $\rho _2$ over $\BB$ and, in
virtue of $C^*$-independence, consider the state $\rho$ over $\BA
\vee \BB$ such that its marginal states are $\rho _1$ and $\rho
_2$ respectively. Then $T^{*}\rho$ is also a state over
$\BA\vee\BB$. If its restriction $(T^{*}\rho) |_{\BB}$ is equal to
$\rho _2$, then $T$ is said to convey no information to
Bob.\end{itemize}

With the original formulation of Definition~\ref{defnoinff}, proof
of Equation~\ref{cninf} is problematic. We show how to prove this
equation with each of the three alternative definitions. First
observe the following remark.
\begin{rem}
Each $C^*$-algebra has sufficient states to discriminate between
any two observables (i.e., if $\rho(A)=\rho(B)$ for all states
$\rho$, then $A=B$). \label{reminff}\end{rem} \noindent To justify
(\ref{cninf}), the CBH authors then say:
\begin{quote}
$(T^*\rho)|_{\BB}=\rho |_{\BB}$ if and only if
$\rho(T(B))=\rho(B)$ for all $B\in\BB$ and for all states $\rho$
on $\BA\vee\BB$. Since all states of $\BB$ are restrictions of
states on $\BA\vee\BB$, it follows that $(T^*\rho)|_{\BB}=\rho
|_{\BB}$ if and only if $\omega(T(B))=\omega(B)$ for all states
$\omega$ of $\BB$, i.e., if and only if $T(B)=B$ for all $B\in
\BB$.
\end{quote}
Let us examine this derivation under each of the three alternative
definitions of conveying no information. By the definition of
$T^*$, we have $(T^*\rho)(B)=\rho(T(B))$ for all states $\rho$
over $\BA\vee\BB$. To obtain from this that $\rho(T(B))=\rho(B)$,
one must show that $(T^*\rho)(B)=\rho(B)$, and this is equivalent
to saying that $(T^*\rho)|_{\BB}=\rho |_{\BB}$ for all states
$\rho$ over $\BA\vee\BB$. Now, according to CBH, one would need to
show that $\rho(T(B))=\rho(B)$ if and only if
$\omega(T(B))=\omega(B)$ with states $\rho$ over $\BA\vee\BB$ and
$\omega$ over $\BB$. The latter formula, however, is not
well-defined: operator $T(B)$, generally speaking, is not in
$\BB$. Fortunately, we are salvaged by the first alternative
reformulation of Definition~\ref{defnoinff}: because
$\rho(T(B))=\rho(B)$ is true for all states $\rho$ over
$\BA\vee\BB$, we obtain directly that $T(B)=B$ in virtue of
Remark~\ref{reminff}.

The second alternative definition of conveying no information
makes use of an object such as $(T|_{\BB})^{*}\rho$. To give it a
meaning in the algebra $\BB$, one needs to impose a closure
condition on the action of $T$ on operators $B\in \BB$: namely,
that $T$ must not take operators out of $\BB$. The problem here is
the same as the one we encountered in the discussion of the
previous alternative, and it is only by assuming the closure
condition that one is able to obtain that $T(B)=B$.

In the third alternative, for the state $\rho$ over $\BA\vee\BB$,
write from the definition of $T^*$ that $(T^*\rho)(B)=\rho(T(B))$.
The result $(T^*\rho)(B)$ is the same as $(T^*\rho)|_{\BB}(B)$,
and this is equal to $\rho _2 (B)$. Consequently, $\rho(T(B))=\rho
_2(B)=\rho (B)$. Can we now say that this holds for all states
$\rho$ over $\BA\vee\BB\;$? The answer is obviously yes, and this
is because each state over $\BA\vee\BB$ can be seen as an
extension of its own restriction to $\BB$. Therefore, one has to
modify Definition~\ref{defnoinff} for it to be formally correct,
and this entails a modification in the proof of
Equation~\ref{cninf}.

We now turn to the remaning two CBH axioms. Axiom CBH2 is used to
establish that algebras $\BA$ and $\BB$, taken separately, are
non-Abelian. Broadcasting, which enters in the formulation of the
axiom, is defined as follows:

\begin{defn}Given two isomorphic, kinematically independent $C^*$-algebras
$\BA$ and $\BB$, a pair $\{\rho _1,\rho _2\}$ of states over $\BA$
can be broadcast in case there is a standard state $\sigma$ over
$\BB$ and a dynamical evolution represented by an operation $T$ on
$\BA\vee\BB$ such that $T^*(\rho _i\otimes\sigma) |_{\BA}=T^*(\rho
_i\otimes\sigma) |_{\BB}=\rho _i$, for $i=0,1$.
\end{defn}

Equivalence between the `no broadcasting' condition and
non-Abelianness of the $C^*$-algebra is then derived from the
following theorem:

\begin{thm}Let $\BA$ and $\BB$ be two kinematically independent $C^*$-algebras.
Then:\begin{description}\item[(i)] If $\BA$ and $\BB$ are Abelian
then there is an operation $T$ on $\BA\vee\BB$ that broadcasts all
states over $\BA$.\item[(ii)] If for each pair $\{\rho _1,\rho
_2\}$ of states over $\BA$, there is an operation $T$ on
$\BA\vee\BB$ that broadcasts $\{\rho _1,\rho _2\}$, then $\BA$ is
Abelian.
\end{description}\label{nobroadth}\end{thm}

It is interesting to note that non-Abelianness of the algebras
$\BA$ and $\BB$, taken one by one, is proved by assuming that they
are kinematically independent. This means that quantumness, of
which non-Abelianness is a necessary ingredient, is not a property
of any given system taken separately, as if it were the only
physical system in the Universe; on the contrary, to be able to
derive the quantum character of the theory, one must consider the
system in the context of at least one other system that is
physically distinct from the first one. As a consequence, for
example, this forbids treating the whole Universe as a quantum
system if one reconstructs quantum theory along the CBH lines.

Axiom CBH3 entails nonlocality: spacelike separated systems must
at least sometimes occupy entangled states. In particular, CBH
show that if Alice and Bob have spacelike separated quantum
systems, but cannot prepare any entangled state, then Alice and
Bob can devise an unconditionally secure bit commitment protocol.
The derivation starts by showing that quantum systems are
characterized by the existence of non-uniquely decomposable mixed
states: a $C^*$-algebra $\BA$ is non-Abelian if and only if there
are distinct pure states $\omega _{1,2}$ and $\omega _{\pm}$ over
$\BA$ such that $\frac{1}{2}(\omega _1+\omega
_2)=\frac{1}{2}(\omega _+ +\omega _-)$. This result is used to
prove a theorem showing that a certain proposed bit commitment
protocol is secure if Alice and Bob have access only to
classically correlated states (i.e. convex combinations of product
states).

\begin{thm}[the CBH `no bit commitment' theorem]
If $\BA$ and $\BB$ are non-Abelian then there is a pair $\{\rho
_0,\rho _1\}$ of states over $\BA\vee\BB$ such that:
\begin{enumerate}
    \item $\rho _0 |_{\BB}=\rho _0 |_{\BB}$.
    \item There is no classically correlated state $\sigma$ over
    $\BA\vee\BB$ and operations $T_0$ and $T_1$ performable by
    Alice such that $T^*_0\sigma=\rho _0$ and $T^*_1\sigma=\rho
    _1$.
\end{enumerate}
\label{nobitcommth}\end{thm}

From this theorem the authors deduce that the impossibility of
unconditionally secure bit commitment entails that ``if each of
the pair of \textit{separated} physical systems $\BA$ and $\BB$
has a non-uniquely decomposable mixed state, so that $\BA\vee\BB$
has a pair $\{\rho _0, \rho _1\}$ of distinct classically
correlated states whose marginals relative to $\BA$ and $\BB$ are
identical, then $\BA$ and $\BB$ must be able to occupy an
entangled state that can be transformed to $\rho _0$ or $\rho _1$
at will by a local operation.'' The term `separated' is essential
and, nevertheless, its precise meaning is not defined. It can be,
indeed, compared to the use of terms `distinct' and `distant' in
the analysis of Axiom CBH1. When the authors claim that Alice and
Bob represent ``\textit{spacelike} separated systems,'' while
formally Alice and Bob are just two $C^*$-algebras, one sees how
the way in which CBH apply the algebraic formalism is severely
constrained by the context of quantum information theory. Here
appears again a situation in which language and context used to
formulate and justify the fundamental principles set up a limit on
the applicability of the mathematical formalism in which these
principles are then represented. Even if the formalism can be
understood more generally than within the discipline chosen in
order to comprehend the language, one still cannot make his way
out of this disciplinary prison or else the sense of the axioms
will be lost. If one however persists and crosses the border and
then, say, obtains a new mathematical result, this result will be
void of physical meaning until a new, broader justification of the
fundamental principles is given. Philosophical and linguistic
justification, and mathematical derivation play here a game of
mutual onslaught and retreat which, ultimately, leads to the
advance of science.

The CBH result would be a perfect example of reconstruction were it
not for a great deal of mathematical structure which is tacitly
assumed in the choice of $C^*$-algebra as a mathematical
representative of the notion of system. Assumptions of the algebraic
formalism include the relations between operators abiding by the
linear law, numeric coefficients in algebras being complex numbers,
the states giving rise to the Hilbert space representation via the
GNS construction, etc. Once one lists all these supplementary
assumptions, the CBH reconstruction appears once again to suffer
from a similar defect of incorporating a serious mathematical
abstraction as derivations from axioms H1-H5 or R1-R2.

\section{Conclusion}

Reconstruction brings in clarity to where interpretation was
struggling to make sense of a physical theory. What belongs to
physical theory is no more than what is needed for its derivation.
All other questions belong to metatheory and are related to the
metatheoretic justification task for the choice of first
principles.

The notion of reconstruction presented here resembles Einstein's
notion of `principle theory'. Principle theories, according to
Einstein, ``employ the analytic, not the synthetic, method. The
elements which form their basis and starting point are not
hypothetically constructed but empirically discovered ones,
general characteristics of natural processes, principles that give
rise to mathematically formulated criteria which the separate
processes or the theoretical representations of them have to
satisfy''~\cite{Einstein1954}. One recognizes here what we have
called justification of first principles and a statement that the
mathematical derivation must follow after the principles had been
established. Einstein's distinction between constructive and
principle theories, though, bears a heavy flavor of his
ontological view of relativity theory. For quantum theory, as Bub
first realized~\cite{bubPT}, Einstein's notion of principle theory
is still applicable although, as we argued above, with a modified
status of the first principles.

Reconstruction of quantum theory remains an only partially solved
problem. Notwithstanding, it is already competing with traditional
interpretations due to its appealing conceptual transparency and to
the clarity that it brings into the structure of the theory. It
would be too ambitious to expect that all of modern quantum theory,
including field theory and quantum gravity, could be derived from a
few axioms; mathematical abstractions and further assumptions are
still a necessity. However, if we want to understand the meaning of
even most advanced parts of quantum theory, it is inevitable that
simple physical principles be formulated and put in the very
foundation of quantum theory.

\singlespacing

\end{document}